\documentclass[a4paper,UKenglish,cleveref, autoref, thm-restate]{lipics-v2021}

\usepackage{amsmath}  

\newcommand{\ignore}[1]{}

\newcommand{\universe}{\mathcal{U}}
\newcommand{\eps}{\varepsilon}
\newcommand{\costsigma}{\mbox{cost}^{\text{\footnotesize MIX}}}
\newcommand{\costmu}{\mbox{cost}^{\text{\footnotesize LA}}}
\newcommand{\costbeta}{\mbox{cost}^{\text{\footnotesize EXP}}}

\title{Blocked Bloom Filters with Choices}

\author{Johanna Elena Schmitz\footnote{\label{contribution}Contributed equally.}}{Algorithmic Bioinformatics, Faculty of Mathematics and Computer Science, Saarland University \and Saarbrücken Graduate School of Computer Science \and Center for Bioinformatics Saar, Saarland Informatics Campus, Saarbrücken, Germany}{jschmitz@cs.uni-saarland.de}{https://orcid.org/0009-0002-6377-2561}{}  

\author{Jens Zentgraf\,$^1$}{Algorithmic Bioinformatics, Faculty of Mathematics and Computer Science, Saarland University \and Saarbrücken Graduate School of Computer Science \and Center for Bioinformatics Saar, Saarland Informatics Campus, Saarbrücken, Germany}{zentgraf@cs.uni-saarland.de}{https://orcid.org/0000-0001-9444-2755}{}

\author{Sven Rahmann\footnote{Corresponding author: rahmann@cs.uni-saarland.de}}{Algorithmic Bioinformatics, Faculty of Mathematics and Computer Science, Saarland University \and Center for Bioinformatics Saar, Saarland Informatics Campus, Saarbrücken, Germany}{rahmann@cs.uni-saarland.de}{0000-0002-8536-6065}{}

\authorrunning{J. E. Schmitz, J. Zentgraf and S. Rahmann}  

\Copyright{Jane Open Access and Joan R. Public}  



\category{} 

\relatedversion{} 

\nolinenumbers 

\hideLIPIcs

\EventEditors{John Q. Open and Joan R. Access}
\EventNoEds{2}
\EventLongTitle{42nd Conference on Very Important Topics (CVIT 2016)}
\EventShortTitle{CVIT 2016}
\EventAcronym{CVIT}
\EventYear{2016}
\EventDate{December 24--27, 2016}
\EventLocation{Little Whinging, United Kingdom}
\EventLogo{}
\SeriesVolume{42}
\ArticleNo{23}

\begin{document}

\maketitle
\begin{abstract}
Probabilistic filters are approximate set membership data structures that represent a set of keys in small space, and answer set membership queries without false negative answers, but with a certain allowed false positive probability.
Such filters are widely used in database systems, networks, storage systems and in biological sequence analysis because of their fast query times and low space requirements.
Starting with Bloom filters in the 1970s, many filter data structures have been developed, each with its own advantages and disadvantages, e.g., Blocked Bloom filters, Cuckoo filters, XOR filters, Ribbon filters, and more.

We introduce \emph{Blocked Bloom filters with choices} that work similarly to Blocked Bloom filters, except that for each key there are two (or more) alternative choices of blocks where the key's information may be stored. When inserting a key, we select the block using a cost function which takes into account the current load and the additional number of bits to be set in the candidate blocks.
The result is a filter that partially inherits the advantages of a Blocked Bloom filter, such as the ability to insert keys rapidly online or the ability to slightly overload the filter with only a small penalty to the false positive rate.
At the same time, it avoids the major disadvantage of a Blocked Bloom filter, namely the larger space consumption. 
Our new data structure uses less space at the same false positive rate, or has a lower false positive rate at the same space consumption as a Blocked Bloom filter.
We discuss the methodology, cost functions for block selection, engineered implementation, a detailed performance evaluation and use cases in bioinformatics of Blocked Bloom filters with choices, showing that they can be of practical value.

The implementation of the evaluated filters and the workflows used are provided via Gitlab at \url{https://gitlab.com/rahmannlab/blowchoc-filters}.

\end{abstract}

\section{Introduction}
\label{sec:intro}

\subparagraph{Probabilistic filters.}
A probabilistic filter, in the sense used here, is an approximate set membership data structure that in its most basic form supports only two operations:
inserting a key, and querying a key, i.e., asking whether it is contained in the represented set, with the following semantics:
After inserting a key, the containment query for it always returns \texttt{True}. 
However, even if a key was never inserted into the filter, its containment query may nonetheless falsely return \texttt{True} with a small controllable probability, called the false positive rate (FPR)~$\eps$.
The space requirement of a filter is independent of the key size and only depends on the FPR.
This is achieved by storing a small fingerprint or other kind of bit pattern instead of the key.
In order to store $n$~keys with an FPR of $\eps = 2^{-k}$, at least $nk$ bits are required, but practical filters need $Cnk$ bits with an \emph{overhead} factor $C>1$.


\subparagraph{Applications of filters.}
Filters typically have relatively simple implementations, can be queried quickly and need moderate space in comparison to an exact representation.
Often, they can be kept entirely in memory, even if the complete exact representation cannot.
This makes them useful in practice, e.g., in connection with databases, as a pre-filtering tool: Only if the filter says that a queried key may be in the database, an (expensive) database query needs to be launched.
Our main interest lies in genomics, and more precisely in the analysis of very large amounts of DNA sequence data.
Concrete use cases for the new filter types introduced in this work are presented in the evaluation and discussion sections.

\subparagraph{Existing types of probabilistic filters.}
Many different variations of filters exist, often with additional capabilities, such as the possibility to delete keys (under specific conditions), to count how many times the same key was inserted (up to a limit), or to store additional data with keys.
We discuss some filter types below.

Bloom filters \cite{bloom_spacetime_1970} are the most well known classical example of filters. 
Their overhead factor is relatively high with $C = 1/\ln(2)\approx 1.443$.
In addition, for $\eps = 2^{-k}$, Bloom filters access $k$ (deterministically computed) randomly distributed bits in memory to insert or query one key.
Each of these accesses is a likely cache miss; so insertion and lookup speeds decrease with lower FPRs.

These disadvantages led to the development of other types of filters, including several variants of the Bloom filter, such as the Blocked Bloom filter \cite{putze_cache-_2007}, and filters based on different principles, such as the Cuckoo filter \cite{fan_Cuckoo_2014}, (Vector) Quotient filter \cite{bender_dont_2012,pandey_vector_2021}, XOR filter \cite{graf_xor_2020}, Binary Fuse filter \cite{graf_binary_2022}, or (Bumped) Ribbon filter \cite{dillinger_ribbon_2021,dillinger_fast_2022}, to mention but a few.
The goals were to reduce the space requirement (for the same FPR), or to improve insert or lookup speed, or all of these simultaneously.
Another desirable property of a filter is extendability, i.e., its dynamic growth as more elements are inserted.
The filters mentioned above are not extendable and have to be initialized with a given storage capacity, but more recent filters, such as Aleph filters \cite{dayan_aleph_2024} or Infini filters  \cite{dayan_infinifilter_2023} are extendable, but come with different disadvantages.

A considerable advantage of Bloom, Cuckoo and (Vector) Quotient filters is that keys can be inserted online as they appear, i.e., it is not necessary to know the entire set of keys beforehand, which is necessary for other filters (XOR, Binary Fuse, Ribbon filters).
Among the filters with online insertion capabilities (in a recent tutorial and review \cite{pandey_beyond_2024}, they are called semi-static), Bloom filters and their relatives stand out because they can be filled over capacity, in which case they do not immediately fail, but continue to work, albeit with a higher FPR.
This is in contrast to Cuckoo and Quotient filters, which cannot be filled over capacity unless one also allows false negative queries.
In addition, Cuckoo and Quotient filters may move data of already inserted keys around as more keys are inserted, which can lead to long and inhomogeneous insertion times when the filter is close to full, whereas insertion time for a Bloom filter is constantly $\Theta(k)$.

The advantage of a Blocked Bloom filter over a standard Bloom filter is that the $k$~accessed bits are not randomly distributed in memory, but concentrated in one block, which ideally corresponds to a single cache line (of typically 512 bits).
This change reduces the number of cache misses from~$k$ to~$1$ for each insertion or lookup operation, but comes at the cost of a higher overall FPR because the load (fraction of 1-bits) per block is heterogeneous; see~Section~\ref{sec:background} for an explanation and details.

\begin{figure}[t]
\centering
\begin{minipage}{0.65\textwidth}
\scalebox{0.9}{
\addtolength{\tabcolsep}{-0.2em}
$\!\!\!$\begin{tabular}{llccc}
            & Best-case  & Worst & \multicolumn{2}{c}{Cache misses for} \\ 
Name [Ref.] & overhead factor $C(k)$ & $C(10)$ & insert & lookup \\
\hline
Bloom \cite{bloom_spacetime_1970} & 1.443 & 1.443 & $k$ & $k$  \\
Blocked Bloom \cite{putze_cache-_2007} & $1.443 + F(k)$ ${}^{(\dagger)}$ & 1.565 & 1 & 1   \\
BlowChoc [this] & 1.443 + $f(k)$ ${}^{(\dagger)}$ & 1.456 & $2$ & 1 to $2$ \\
\hline  
Cuckoo \cite{fan_Cuckoo_2014} & $\ge 1.053 \,(1+3/k)$ ${}^{(\ddagger)}$ & 2.73 & 1 to $M{}^{(*)}$ & 2\\
Quotient \cite{bender_dont_2012} & $\ge 1.053\,(1+2.125/k)$ ${}^{(\ddagger)}$ & 2.55 & some${}^{(**)}$ & some${}^{(**)}$ \\
Vector quotient \cite{pandey_vector_2021} & $\ge 1.0753\,(1+2.914/k)$ ${}^{(\ddagger)}$ & 2.77 & 2 & 2\\
\end{tabular}}
\end{minipage}
\hfill
\begin{minipage}{0.25\textwidth}
\centering
\includegraphics[width=1.0\linewidth]{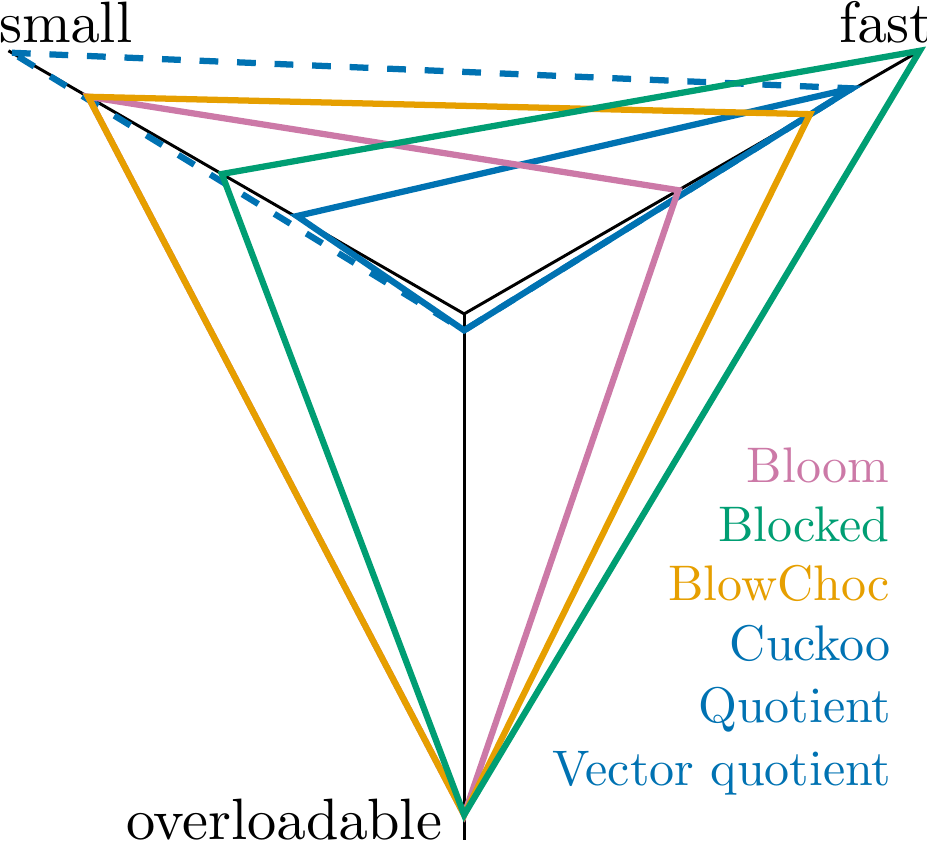}
\end{minipage}
\caption{
Overview of filter types supporting online insertions.
Only the Bloom types (above the middle separator in the table) support smooth FPR degradation when the filter is overloaded.
The overhead factor $C=C(k) > 1$ specifies the required space per key; one needs $Cnk$ bits to store $n$ keys with an FPR of $2^{-k}$.
${}^{(\dagger)}$: $F$ and $f$ are increasing functions of $k$ with no known closed form; $F$~increases faster than~$f$.
${}^{(\ddagger)}$: The number of buckets/slots in the filter must be a power of~2. Therefore, the overhead factor can be up to twice the lower bound given here; we provide the worst case for $k=10$.
The number of cache misses is a proxy of insertion/lookup time.
${}^{(*)}$:~For Cuckoo filters, $M$ is the maximum random walk length during insertion. Higher values of~$M$ achieve lower overhead but require more time.
${}^{(**)}$:~For Quotient filters, cache misses depend on the load factor; higher loads (lower overhead) means more cache misses and more time.
The triangles on the right hand side qualitatively illustrate the filter properties concerning space usage, build/query time and overloadability.
}\label{fig:filtertypes}
\end{figure}

A summarizing comparison of the filters allowing online insertion is shown in Figure~\ref{fig:filtertypes}.
The space overhead required by the Bloom filter family is tolerable, whereas the overhead of the other filters can be both lower and considerably higher, depending on circumstances.
Therefore, and because of the possibility to overload them with smooth FPR degradation, we are primarily interested in the Bloom filter family.
It is natural to ask whether the simplicity and speed of Blocked Bloom filters can be combined with other techniques to achieve a lower FPR or smaller space overhead, while retaining the smooth FPR degradation properties of Bloom filters when more keys are inserted than initially planned.

\subparagraph{The power of choices.}
We leverage the ``power of (two or more) choices'' idea \cite{mitzenmacher_power_2001} to unite the advantages of Blocked and standard Bloom filters (fast speed, little extra overhead).
The idea to use choices within Bloom filters is not entirely new.
It was already explored for standard (not Blocked) Bloom filters, using two independent sets of bit positions in the entire bit array \cite{lumetta_using_2007}.
This approach needs $2k$ memory accesses with cache misses instead of the already large $k$, further reducing throughput. 
The experimental results also showed no distinctive other advantages.
In another approach, balanced Blocked Bloom filters \cite{kanizo_access-efficient_2013}, a key is placed in the first block which stores less than a pre-defined number of keys. 
A counter monitors the number of keys in each block.
Counting keys is faster but less informative than looking at actual bit usage, and the approach led to only small improvements.

\subparagraph{Contributions.}
Our main contribution is to use the structure of a Blocked Bloom filter, but choose between two blocks (given by two distinct block address hash functions) based on a new engineered cost function.
In this way, we obtain a more balanced block load, and we can optimize the placement of keys to re-use already used bits, together leading to a lower load overall and therefore a lower FPR.
It will be shown in extensive evaluations that the effects of optimizing the key placements using a cost function are strong enough to more than counteract the FPR doubling caused by searching at two different locations.
This is in contrast to balanced Blocked Bloom filters, which do not optimize the key placement and therefore do not efficiently re-use already set bits, resulting in a filter with a higher FPR compared to Bloom filters \cite{kanizo_access-efficient_2013}. 

After providing technical background on Bloom and Blocked Bloom filters in Section~\ref{sec:background}, we introduce Blocked Bloom filters with choices (BlowChoc filters) in Section~\ref{sec:methods}, where we also discuss an insertion cost function and its parameters.
A comparative evaluation against its closest relatives, the Bloom filter and the Blocked Bloom filter, follows in Section~\ref{sec:results}, together with application cases in genomics.
We conclude with a summarizing discussion. 


\section{Background}
\label{sec:background}


\subsection{Bloom filters}
\label{sec:bloom}

A Bloom filter consists of an array of $m$~bits (initially all zero) and $k$~independent random hash functions from a universal family, $f_1,\dots,f_k: \mathcal{U} \to [m] := \{0, \dots, m-1\}$, mapping the universe~$\mathcal{U}$ of possible keys to bit addresses.
To insert a key $x\in\mathcal{U}$, we set all bits $f_1(x), f_2(x), \dots, f_k(x)$ to~1.
To query the membership status of $x$, we return the logical \texttt{and} (product) of these bits, $\bigwedge_{i=1}^k\, f_i(x)$.
If $x$~is a key that was never inserted, and the load factor (fraction of 1-bits) at query time is~$p_1$, then the probability of a false positive equals the probability that all queried $k$ bits are 1 (set by previous insertions), which is~$p_1^k$.

The size $m$ of the Bloom filter is chosen such that after inserting $n$ keys, the (expected) load is $p_1 = 1/2$, making the bit array incompressible and yielding an FPR of $2^{-k}$.
To compute~$m$, we note that $p_1=1/2$ is equivalent to $p_0 := 1 - p_1 = 1/2$ (fraction of unset bits).
By the law of large numbers (with $n, m$ in the millions or billions), this (random) fraction is tightly concentrated around its expected value, which equals the probability that any fixed bit is not set.
A bit is not set if it was never chosen by all $nk$ bit set operations; so the probability for this event (for large $m,n,k$) is
\begin{equation}\label{eq:bloom:p0}
    p_0 := (1-1/m)^{nk} \approx \mbox{e}^{-nk/m} \,.
\end{equation}
Solving $p_0 = 1/2$ for~$m$ yields $m = nk/\ln(2) \approx 1.443\, nk$; hence the well-known space overhead factor of $C\approx 1.443$, independently of~$k$.

The assumption behind the calculations above is that the $k$~hash functions behave as if they select $k$ distinct but otherwise independent random bit addresses.
In practice, one can simply choose $k$ random functions from a sufficiently large family of simple hash functions.
In very rare cases, two (or more) bit addresses may coincide, such that we effectively set or query less than $k$ distinct bits for a very small number of keys; this effect is so small that it has no measurable influence on the FPR when $m$~is large and $k$ is a small constant.
However, this consideration can be important for Blocked Bloom filters.

A Bloom filter may be overloaded without failing completely.
In that case, its FPR increases beyond the desired value.
The expected fraction $p_1 = 1 - p_0$ of set bits can be computed from Eq.~(\ref{eq:bloom:p0}) by replacing $n$ by $\gamma n$ with the appropriate overloading factor $\gamma > 1$, yielding $p_1 = 1 - p_0^\gamma = 1 - 2^{-\gamma}$, and the resulting FPR is concentrated around 
\begin{equation}\label{eq:bloom:eps}
  \eps(\gamma, k) = (1 - 2^{-\gamma})^k \,,
\end{equation}
showing the smooth dependence of the FPR $\eps$ on the overloading factor $\gamma$.
For example, for 10\% overload ($\gamma=1.1$) and $k=10$, we obtain an FPR of roughly twice the desired one ($1.912\cdot 2^{-10}$ instead of $2^{-10}$).
Conversely, when we query an underfull Bloom filter, the FPR is better (lower) than the desired one.

\subsection{Blocked Bloom filters}
\label{sec:blocked}

The $k$ random memory accesses (cache misses) for each insertion or query into a Bloom filter can lead to low throughput, even though modern optimizing compilers and the prefetching capabilities of current CPUs may counteract these effects to some degree.
The idea behind Blocked Bloom filters \cite{putze_cache-_2007} is to access not $k$ randomly distributed bits, but $k$ bits in close proximity, i.e., in the same block.
Ideally, a block corresponds to a cache line (typically 512 bits; for simplicity of presentation, we will use this constant value throughout this article).
Therefore, a Blocked Bloom filter consists of $m = 512\,M$ bits, divided into $M$ blocks of 512 bits, a \emph{block address} hash function $h_1: \mathcal{U} \to [M]$ and a number $k$ of independent \emph{bit address} hash functions, randomly chosen from a  universal family, $f_1,\dots,f_k: \mathcal{U} \to [512]$.

The obvious advantage of a Blocked Bloom filter over a standard Bloom filter is the increased speed, especially for low FPRs (high values of~$k$).
However, there is also a substantial disadvantage.
For the same size as a  Bloom filter, the overall FPR of a Blocked Bloom filter is higher due to two reasons: 
First, the $M$~blocks have different loads.
Blocks with only few remaining unset bits have a high local FPR, while blocks with many remaining unset bits have a low local FPR.
On average, because of Eq. (\ref{eq:bloom:eps}) and Jensen's inequality, the blocks with high load increase the average FPR more significantly than blocks with low load decrease it.
The surplus FPR caused by inhomogeneity increases with~$k$; see~Section~\ref{sec:results} (Figure~\ref{fig:overhead}) for concrete numbers.
The effect can be compensated by using more bits (hash functions) per key and/or more blocks, thus increasing the space overhead (but maintaining the desired FPR).
Second, when independent random hash functions are chosen, the probability of a bit address collision with $k$ out of 512 can be substantially high, such that in several cases, only $k' < k$ bits are effectively queried, increasing the FPR from $p_1^{-k}$ to $p_1^{-k'}$.
This effect can be compensated by choosing the bit addresses not independently, but in a way that ensures distinct addresses (see~Section~\ref{sec:bitselection}), at the cost of higher insertion and lookup times.

Therefore, while offering a speed advantage over standard Bloom filters, Blocked Bloom filters either yield undesirably higher FPRs or need more space.
We wondered whether it would be possible to design a variant of Bloom filters that offers close to the same speed advantage without incurring these penalties.


\begin{figure}[t]\centering
\includegraphics[width=1.0\linewidth]{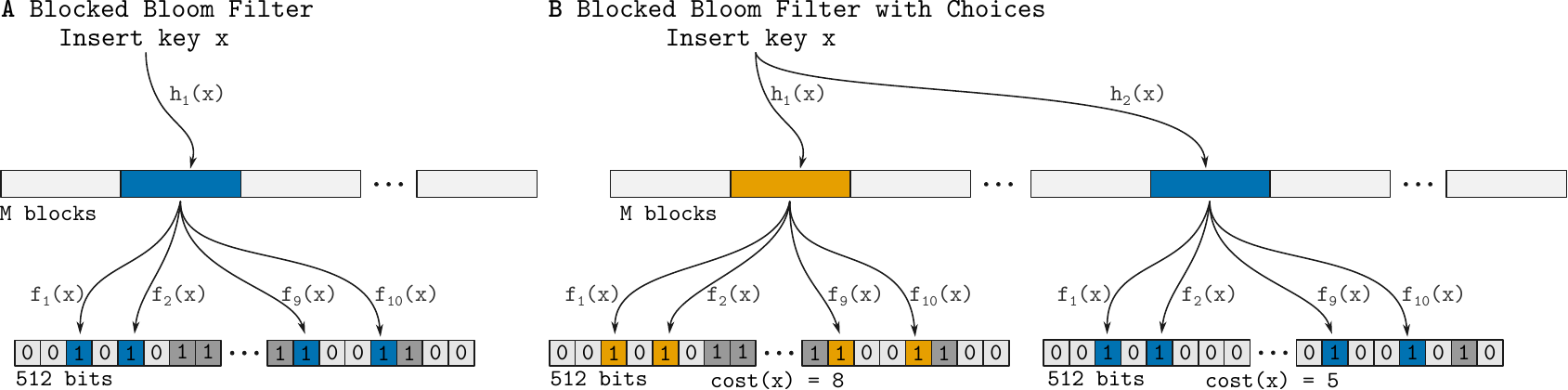}
\caption{
Inserting a key into a Blocked Bloom filter, with and without choices. 
\textbf{(A)} The key is hashed to a single block, and $k$~bits are set in this block.
\textbf{(B)} The key may be inserted into one of two possible blocks. After computing the insertion cost for both blocks, the $k$ bits are set in the block with lower cost.
}\label{fig:bcfilter}
\end{figure}

\section{Blocked Bloom Filter with Choices}
\label{sec:methods}

We introduce the Blocked Bloom filter with choices (``BlowChoc filter'').
After describing the basic idea (Section~\ref{sec:basicidea}), we discuss cost functions for selecting the block to insert the key into among the possible alternatives (Section~\ref{sec:costfunc}), and describe different bit selection strategies within a block (Section~\ref{sec:bitselection}). Implementation details are discussed in Section~\ref{sec:implementation}.

\subsection{Basic idea}
\label{sec:basicidea}

The idea behind the BlowChoc filter is that it works like a Blocked Bloom filter, but there are two (or more) independent choices for the block.
The bits for a key may be set in either block.
When querying the key, both blocks are examined, and the answer \texttt{True} is returned if all queried bits are set in at least one of the blocks. 
In contrast to Cuckoo and Quotient filters, data of inserted keys is never moved after insertion.

At first sight, the idea of BlowChoc filters seems to be disadvantageous: 
By querying two locations, the FPR approximately doubles from $\eps$ to $\eps + \eps - \eps^2 \approx 2\eps$, and accordingly almost triples for three choices.
On the other hand, having a choice between two or more blocks allows us to choose the block where the insertion requires setting fewer new bits, reducing the load factor overall, and balancing the load better between the blocks.
Together, these possibilities can reduce the FPR overall or the required space overhead for the same FPR.
The question is whether in the possible design space there is an insertion strategy that results in an overall advantage.

The main idea and the differences in comparison to a Blocked Bloom filter without choices are illustrated in Figure~\ref{fig:bcfilter}.
As for a Blocked Bloom filter, there are $M$~blocks and $m=512\,M$ bits in total.
There are two \emph{block address} hash functions $h_1, h_2:\mathcal{U}\to [M]$ and $k$~\emph{bit address} hash functions $f_1,\dots,f_k: \mathcal{U}\to [512]$.
From the existing bit pattern in blocks $h_1(x), h_2(x)$ and the bit addresses to be used for key~$x$, given by the set 
\begin{equation}\label{eq:bitset}
    F(x) := \{ f_i(x) \;|\; i=1,\dots,k \} \,,
\end{equation} 
the costs $C_1, C_2$ of setting the bits indexed by $F(x)$ in either block are computed, and the insertion into the block with lower cost is performed.
A~cost function to compute $C_1$ and $C_2$ is described in Section \ref{sec:costfunc}.
Note that $|F(x)| < k$ is possible if some of the bit addresses collide (see Section \ref{sec:bitselection}).

This basic idea can be generalized from two choices to $c\ge 2$ choices, with the disadvantages of increasing cost evaluation time during insertion and increasing query time, especially for unsuccessful lookups, where all choices have to be examined.

\begin{figure}\centering
\includegraphics[width=1.0\linewidth]{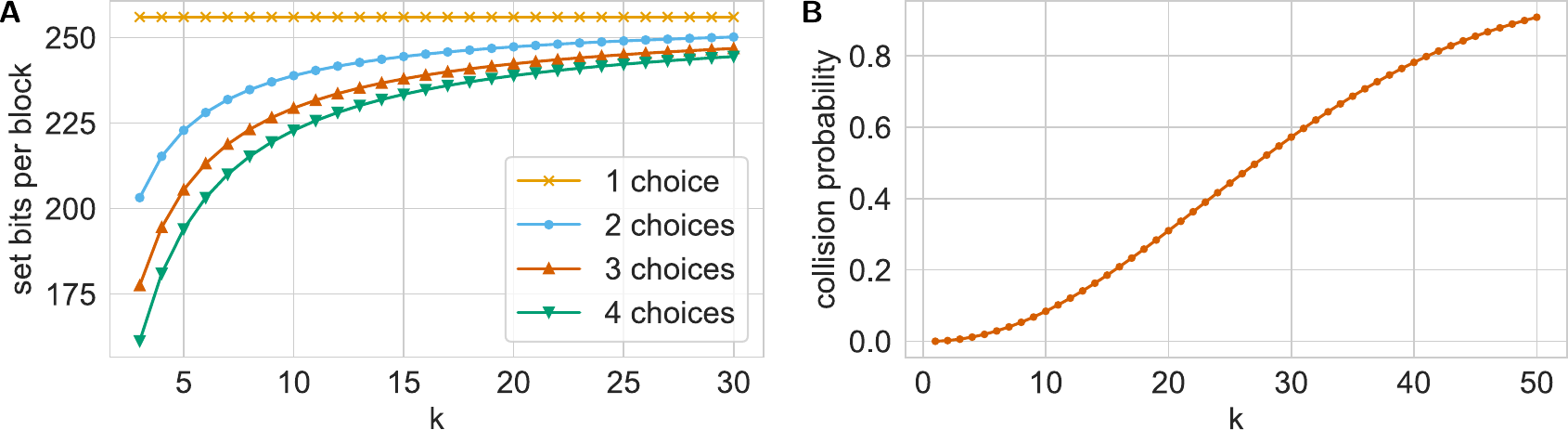}
\vspace{-3ex}
\caption{
\textbf{\sffamily A} Upper bounds for the number of set bits per block to achieve a local FPR $\le 2^{-k} / c$ for $c=1, 2, 3, 4$ and different $k$. 
\textbf{\sffamily B} Probability that $k$ hash values $f_i(x)\in [512]$ from $k$ random hash functions $f_i: \mathcal{U} \to [512]$, $i=1,\dots,k$ have at least one common value, i.e., $|F(x)|<k$ in Eq.~\eqref{eq:bitset}, according to the birthday paradox.
Already for $k=10$, there is an almost 10\% risk of obtaining less than $k$ distinct values.}
\label{fig:maxbits}
\end{figure}

Whether more choices lead to a higher or lower overall FPR depends on whether the multiplied FPR can be compensated for by better load balancing and an overall lower load.
Assuming idealized conditions (distinct bit addresses, homogeneous load in every block), we may ask for the maximally allowed load (number of set bits out of 512) in every block to achieve the desired FPR of $2^{-k}$ if the filter size is the same as for a standard Bloom filter.
Figure~\ref{fig:maxbits}A shows this maximum allowed load for different choices~$c$ (colors) and different $k\ge 3$ (x-axis).
Especially for small~$k$ (a large FPR of $2^{-k}$), the allowed load in each block is considerably smaller for several choices than for Blocked Bloom filters without choices (i.e., one choice).
We therefore need to carefully design a cost function that allows us to achieve both lower loads on average and more balanced loads (ideally constant load per bucket).
It also seems reasonable to expect that the design works better for larger values of~$k$.

\subsection{Cost function for block selection}
\label{sec:costfunc}

When selecting a block, we have two simultaneous goals, which may contradict each other:
On the one hand, we desire a small local FPR (equivalently, a small fraction of set bits within the block).
On the other hand, we desire to set as few additional bits as possible (equivalently, to re-use as many already set bits as possible), keeping the overall load and FPR small.
Therefore, we experimented with different cost functions for selecting the best block for insertion among the choices $h_i(x)$, $i=1,\dots,c$ for a given key~$x$.

First and foremost, if one of the possible blocks already has all bits of index set~$F(x)$ from Eq.~(\ref{eq:bitset}) set to~1, we do not need to change anything, as a query for~$x$ already returns \texttt{True}; so the following considerations do not apply. 

All reasonable cost functions we invented are based on the following properties of a block:
\begin{itemize}
    \item $j$, the \emph{total} number of bits that would be set in the block \emph{after} insertion,
    \item $a$, the number of bits that would have to be newly set in the block.
\end{itemize}
In other words, if $F(x)$ is the set of bit addresses for a key $x$, and $F'$ is the set of already used bits in the block under evaluation, then $j=|F' \cup F(x)|$ and $a=|F(x) \setminus F'| = j - |F'|$.

When comparing two blocks whose load increases beyond $1/4$ (128 of 512 bits set), we penalize more and more steeply the number~$j$ of total set bits in comparison to~$a$.
In an almost empty block, however, parameter~$a$ should be decisive.
After experimentation with different cost function forms (with mixed success; see Appendix~\ref{app:costfun}), we decided to use the following exponential form with a base parameter $\beta>0$:
\begin{equation}
    \costbeta_\beta(j,a) := \beta^{(j / 128)} \,+\,  a / k \,.
    \label{eq:cost:exp}
\end{equation}

We evaluate different values of $\beta$ systematically in Appendix~\ref{app:bestcost}, together with other cost function types and with different bit selection strategies, which we describe next.

\subsection{Bit selection strategies}
\label{sec:bitselection}

We describe alternatives for selecting the set of bit addresses $F$ given by Eq. (\ref{eq:bitset}).
The two strategies that we consider are called \emph{random} and \emph{distinct} and described here.

\subparagraph{Random.} 
The hash functions $f_i$, $i=1,\dots,k$ are chosen randomly and independently from some universal family and are assumed to represent $k$ independent random variables taking uniformly distributed values in the set $[512] = \{0,\dots,511\}$.
The advantage is that the values $f_i(x)$ can be computed efficiently in parallel.
However, if a hash collision occurs, the value set~$F(x)$ from Eq.~(\ref{eq:bitset}) may contain less than $k$ distinct bit addresses (see Figure \ref{fig:maxbits}B).
On the one hand, this decreases the effective number of set bits, yielding an overall lower load factor.
On the other hand, it means that frequently fewer than $k$ bits are queried, leading to a higher overall FPR.
While the effect can be ignored for the standard (non-blocked) Bloom filter, it has a considerable adverse effect on small blocks of 512 bits.

\subparagraph{Distinct.}
Instead of computing $k$ independent random values, we uniformly choose one of the $\binom{512}{k}$ subsets containing exactly~$k$ out of 512~bits.
We use an algorithm based on insertion sort that adapts the bit addresses during insertion. 
For a detailed explanation, see Appendix \ref{app:distinct}.
An evaluation of different combinations of cost functions and bit selection strategies is given in Appendix~\ref{app:bestcost}.

\subsection{Implementation and Optimizations}
\label{sec:implementation}

We have implemented BlowChoc filters with two and more choices and re-implemented Blocked Bloom filters (equivalent to BlowChoc filters with one choice without cost evaluation) and standard Bloom filters. 
Our implementation is available via GitLab (see abstract).
We now discuss several details and optimizations, including parallelization.

\subparagraph{Just-in time compilation using Python with \texttt{numba}.}
The implementation uses just-in-time (JIT) compiled Python based on the \texttt{numba} package \cite{numba}, using typed arrays from \texttt{numpy} \cite{numpy}.
The \texttt{numba} package compiles a subset of Python code into optimized machine code using the LLVM compiler, thereby achieving speeds comparable to that of a C/C++ implementation.
JIT compilation offers additional benefits: Certain parameters can be chosen at Python run time, but before compilation, so they can be compiled into the BlowChoc filter as compile-time constants, allowing for compiler-side optimizations. 
For instance, the array size, the cost function parameter and the parameters of the hash functions can be selected at run time and treated as compile-time constants. 
With custom extensions, we are able to use LLVM intrinsics, like \texttt{popcount} and \texttt{prefetch}, that compile to single CPU instructions.
For example, we use \texttt{popcount} instructions during cost function evaluation to efficiently compute the number of set bits in a block. 
We use software prefetching to preload the blocks for all choices into the L1 cache while evaluating the first choice.

\subparagraph{Parallelization.}
The \texttt{numba} JIT compiler supports efficient parallelization through SIMD vectorization and multithreading.
In order to effectively use $T$ threads in parallel for insertions, the bit array of the filter is subdivided into $T$ sub-arrays of $M/T$ blocks each.
(If necessary, $M$ is rounded up to be a multiple of $T$.)
Each sub-array is cache-line aligned and served by a single thread that writes to it.
The sub-array in which a key is stored is computed by an additional hash function $h_0: \mathcal{U}\to [T]$ that is independent of the block address hash functions $h_1, \dots, h_c: \universe \to [M/T]$. 
A main thread distributes the keys~$x$ to be inserted into the sub-arrays by computing $h_0(x)$ and delegating further work to the appropriate thread via message buffers, thereby avoiding locks on the bit arrays, similarly to previous work \cite{hackgap}.

For query-only workloads, all threads only need read access, and the keys are distributed evenly in chunks across all threads.

\subparagraph{Hash functions.}
The hash functions for addressing the sub-array, the blocks or the bits do not need to satisfy particularly strong independence properties, but they need to be reasonably fast to compute and have sufficiently many free parameters that can be chosen.
In our implementation, we expect $u$-bit unsigned integer keys~$x$ with $u\le 64$ and use linear hash functions of the form $x \mapsto (ax) \bmod b$, where $a$ is a random large integer, and $a$ and $b$ are coprime.  
Note that there is always an implicit mod $2^{64}$ operation after every operation due to the unsigned 64-bit arithmetic; so in fact, we are computing $x \mapsto ((ax) \bmod 2^{64}) \bmod b$.
Modifications for a more uniform distribution of the hash values are applied if $b$~is divisible by~2 or even a power of~2:
If~$b$ is a power of~2, we only consider the $\log_2 b$ most significant bits of $ax$ and use $x\mapsto ((ax) \gg (u - \log_2 b)) \bmod b$.
If $b$ is even but not a power of~2, we perform a bitwise XOR operation between the $q := \lfloor u/2 \rfloor$ most significant bits and the $u/2$ least significant bits of $(ax)$ first, resulting in $x \mapsto ((ax) \oplus ((ax) \gg q)) \bmod b$.

In the insertion implementation, we use an additional array of size~$k$ as working space.
Each bit address value $f_1(x), \dots, f_k(x)$ is evaluated once and stored in the array to evaluate the cost function at all choices (for $c\ge 2$) and to set the bits in the optimal block.


\section{Evaluation}  
\label{sec:results}

We compare the standard Bloom filter, the Blocked Bloom filter and the newly introduced Blocked Bloom filters with 2 and 3 choices, called Blow2Choc and Blow3Choc respectively, along different metrics for different FPRs and numbers of threads.
First (Section~\ref{sec:compare:size}), we consider the resulting FPR for a filter of the same size as a standard Bloom filter, required space for a fixed FPR of $2^{-k}$, 
and the FPR degradation for overloaded filters.
Then (Section~\ref{sec:times}), we compare the throughput for insertions and both successful and unsuccessful lookups.
Finally, we present practical application cases of BlowChoc filters in genomics (Section~\ref{sec:usecase}).

All evaluations were run on a PC workstation with an AMD Ryzen~9 7950X3D CPU with 16~cores and hyperthreading, 128~GB of DDR5 memory with 4800~MHz and CL40 and a Western Digital Black SN850X SSD.
After comparing different cost functions for insertion (Appendix~\ref{app:bestcost}), the $\costbeta_{\beta}$ function with $\beta\approx(1+\sqrt 5)/2$ produced the best overall FPRs.
The appearance of the golden ratio is purely coincidental.
We only consider this particular cost function in the evaluations.

\subsection{FPR and Space Evaluation}
\label{sec:compare:size}

\begin{figure*}\centering
\includegraphics[width=1.0\linewidth]{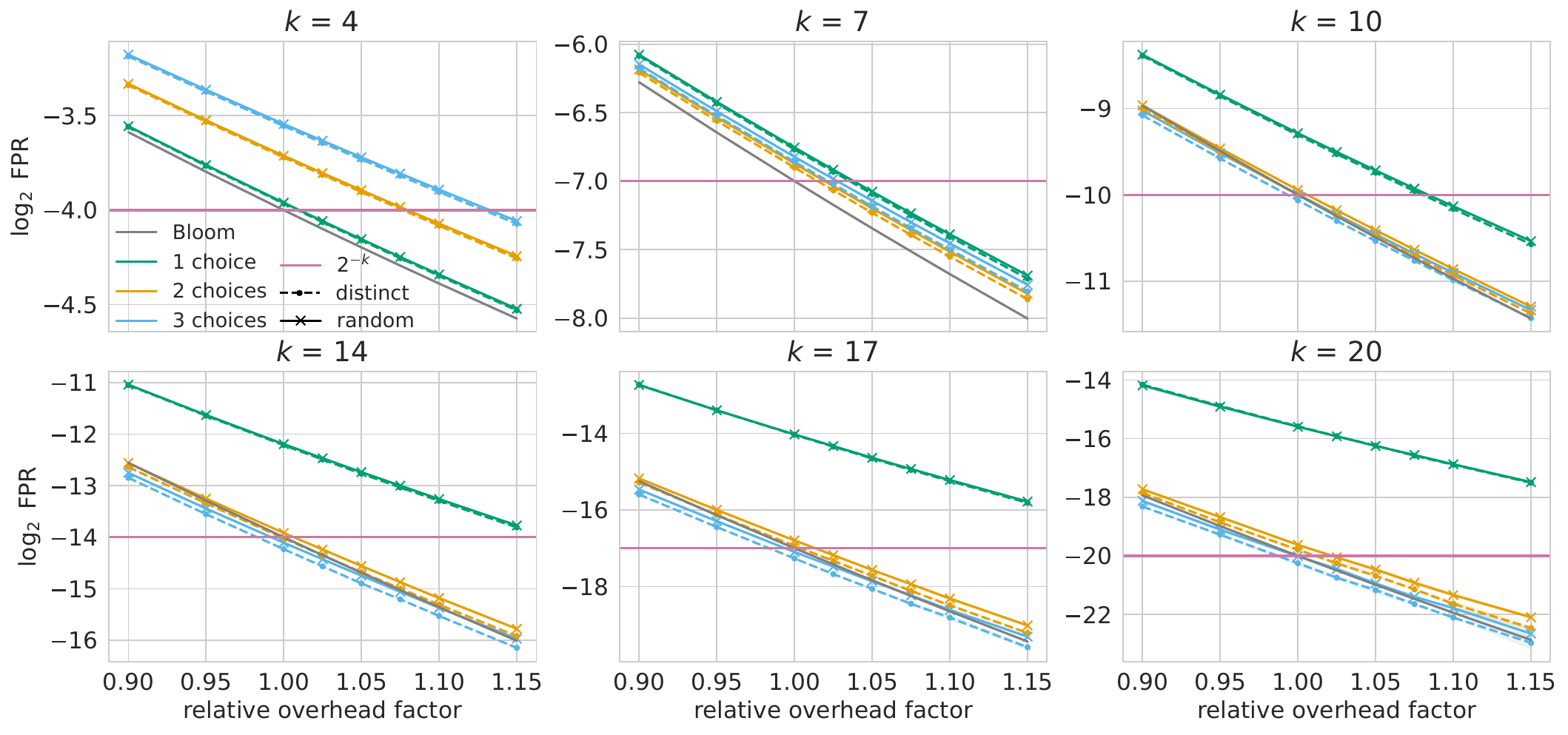}
\caption{
Empirical $\log_2$\,FPR (lower is better) of standard Bloom (grey), Blocked Bloom (1 choice; green) and BlowChoc filters (2 [orange] or 3 [cyan] choices) for the random (solid lines) and distinct (dashed lines) bit selection strategies, using the $\costbeta_\beta$ function with $\beta=(1+\sqrt 5)/2$) for different relative filter sizes (x-axis: relative space overhead) compared to the standard Bloom filter size (at 1.00).
The constant pink horizontal line ($2^{-k}$) shows the targeted FPR of $2^{-k}$.
}\label{fig:overhead}    
\end{figure*}

We created datasets of $2 \cdot 10^9$ random 64-bit integers.
To compute the empirical FPR, we query $N = 2\cdot 10^9$ keys that were not inserted into the filter and count how often the filter erroneously returns \texttt{True} ($W$ times).
The estimated FPR is then $W/N$.
The selected number $N$ allows reasonably accurate empirical evaluations up to $k=20$.
While the FPR could also be computed theoretically from the load histogram of the blocks, the empirical approach has the advantage that it also captures potential problems with the hash functions or with randomization.

\subparagraph{FPR comparison for fixed size.}
Figure~\ref{fig:overhead} shows how the resulting $\log_2$ FPR (which should be $-k$ if $k$ distinct bit address functions are used; marked by pink horizontal lines) changes with the different filter types (colors) and bit selection strategies (line styles) when the filter size is increased above or decreased below the size of a standard Bloom filter (defined as a relative size of $1.00$).
Obviously, the FPR curve for the standard Bloom filter (grey) is always at exactly $-k$ at the relative size overhead factor of $1$ (on the  x-axis) by construction.
A~number of interesting effects can be seen for the other filter types.

As reported previously \cite{putze_cache-_2007}, the standard Bloom filter (grey curve) always has a better (lower) FPR than the Blocked Bloom filter (green curve).
For $k=17$, the Blocked Bloom Filter has an FPR penalty of a factor of~8, or 3 on the $\log_2$ scale.

For small~$k$ (clearly seen for $k=4$), introducing 2 or 3 choices is not beneficial: The FPR curves for 2 and 3 choices are distinctly above those of standard and Blocked Bloom filters, which show comparable performance in this scenario.
There is also almost no observable difference between the random and the distinct bit selection strategy for small~$k$.

For larger~$k$, we observe that the FPR curves for 2 and 3 choices approach that of a standard Bloom filter, whereas the FPR curve for a Blocked Bloom filter becomes worse.
For $k=10$, standard Bloom filters and BlowChoc filters with 2 or 3 choices are comparable in terms of FPR, whereas Blocked Bloom filters are considerably worse by a factor of approximately $2^{0.8}\approx 1.74$.
For $k\ge 14$, we see that 3 choices start to outperform 2 choices and even the standard Bloom filters, and there is a more clearly visible advantage of the distinct bit selection strategy.

These results show that, at sufficiently low FPRs with $k\ge 10$, using choices can indeed be beneficial, and the disadvantages of Blocked Bloom filters are removed by the introduction of choices.
Of course, this has to be considered against a possible increase in computation time (cost function evaluation during insertion; checking two locations during lookup); see Section~\ref{sec:times}.

\subparagraph{Required size for desired FPR of \boldmath$2^{-k}$\unboldmath.}
The typical use case of a filter is to achieve a given FPR, and make the filter as large (or small) as required.
In Figure~\ref{fig:overhead}, we can also see the required relative filter size to achieve the desired FPR of $2^{-k}$: 
We have to find the intersection of the corresponding FPR curve with the pink horizontal line that indicates the targeted FPR of $2^{-k}$.
For Blocked Bloom filters (green), the required relative overhead strongly increases with increasing $k$ and is larger than $1.15$ for $k\ge 14$ and much larger for $k\ge 20$.
In contrast, the overhead factor for BlowChoc filters is largest for $k=4$ and then decreases for increasing~$k$.
For $k \ge 17$, the overhead factor slightly increases again.
For $k \in \{10, \dots, 20\}$, the required relative overhead is slightly above $1.0$ for two choices and $\approx 0.98$ for three choices. 
Hence, BlowChoc filters have similar memory requirements as standard Bloom filters, and can be even smaller. They are much more space efficient than Blocked Bloom filters for $k\ge 10$.

\subparagraph{FPR degradation of overloaded filters.}
From Figure~\ref{fig:overhead}, we can also see that the FPR of all variants of Bloom filters degenerates smoothly if the filter is overloaded by a factor of $\gamma>1$, or equivalently, is initially sized too small by a factor of $\rho := 1/(1+\gamma) < 1$, which corresponds to the size overhead factor in Figure~\ref{fig:overhead}.
Overloading by 10\% is typically feasible without a too severe hit to the FPR, but the penalty increases with~$k$.
The relative penalty (slope of the curves) is similar for all BlowChoc filter variants.
This smooth degradation contrasts the Bloom filter family from other filter types that fail or start reporting false negatives when more keys are inserted than was provisioned for.

\subsection{Throughput Evaluation}
\label{sec:times}

\begin{figure*}[tb]\centering
\includegraphics[width=1.0\linewidth]{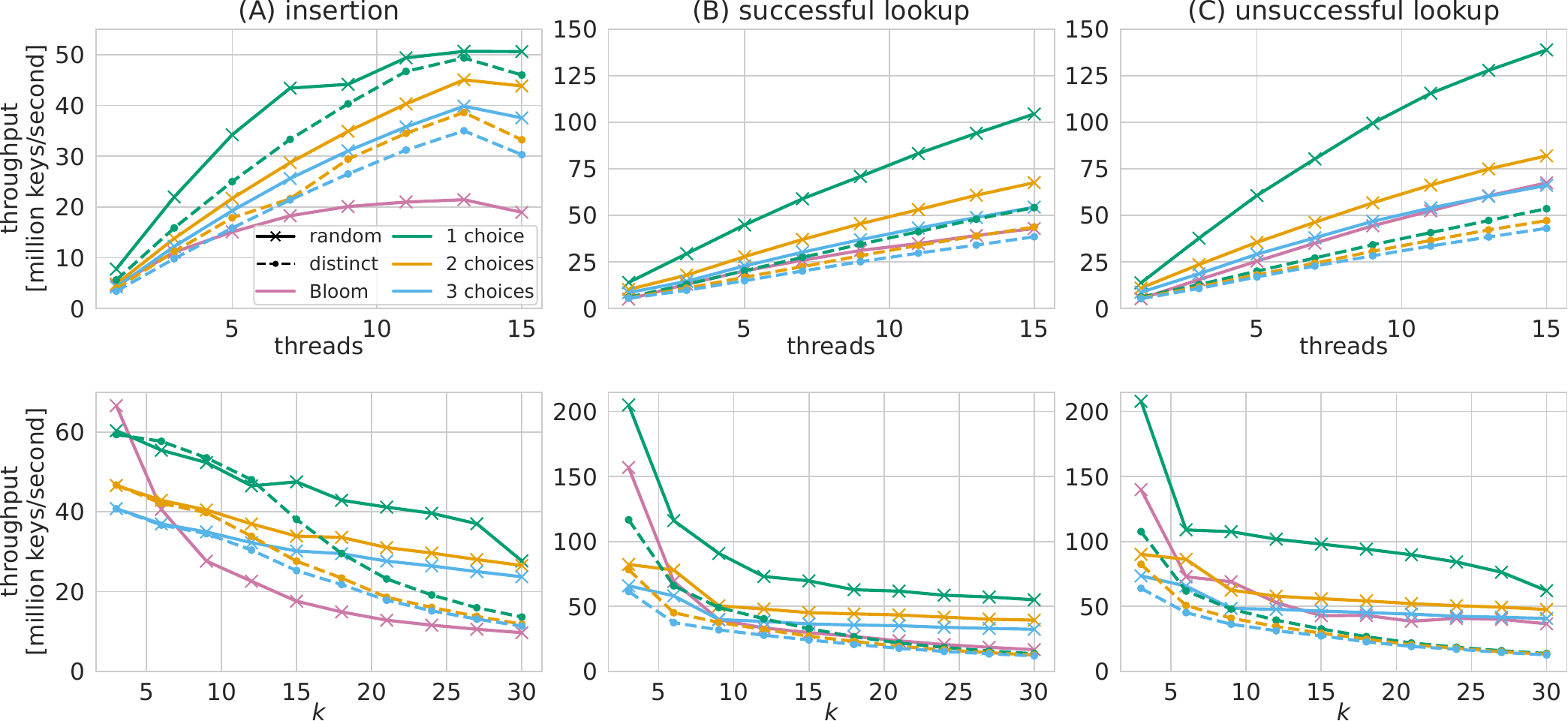}\vspace*{-1ex}
\caption{\normalfont 
Throughput (million keys per second of wall time; higher is better) for (A) inserting, (B) successfully querying, and (C) unsuccessfully querying in a filter with 16 Gbits (2~GB of RAM).
Top row: Throughput for different numbers of subfilters (inserter threads) or query threads, using a fixed number of $k=14$ bit address hash functions.
Bottom row: Throughput for different numbers~$k$ of bit address hash functions, using a fixed number of 9 inserter or query threads.
}\label{fig:times}
\end{figure*}

To benchmark insertion and query throughput, we created different datasets, each containing random 64-bit integers. We configured all filters to have a size of 2~GB ($m=16$ Gbits) and inserted $n=\lfloor m \cdot \ln(2) / k \rfloor$ keys to build a filter with a target FPR of $2^{-k}$.
We measured insertion times, successful lookup times (for previously inserted keys) and unsuccessful lookup times for Bloom, Blocked Bloom and BlowChoc filters with 2 and 3 choices and both bit selection strategies.

Figure~\ref{fig:times} shows the throughput in million keys per second (wall time) for different numbers of inserter threads and different values of~$k$ using the mean of three runs. 
As expected, the throughput of all filters increases in principle with the number of threads. 
Note that, for insertions, the number of threads shown corresponds to the number of subfilters, but we use one additional thread for file reading, and one for distributing the keys to the inserter threads; hence 15 inserter threads actually means 17 running threads, one more than cores on the CPU, explaining the slight decrease in throughput above 14 inserter threads.

As also expected, insertions (column~A in Figure~\ref{fig:times}) are generally slower than queries (columns B and C):
If the key is not already contained, one of the blocks (cache lines) has to be re-written. 
In addition, BlowChoc filters must evaluate the cost function for all blocks during insertion. 

The distinct bit selection strategy is comparably fast to the random strategy for smaller~$k$ but considerably slower for large enough~$k$ (bottom row of Figure~\ref{fig:times}).
This difference is caused by the additional computational overhead of ensuring distinct bits, which includes an implicit sorting step. 
Since the cost for the sorting step increases with~$k$, the gap widens notably for increasing~$k$.
Given the small FPR differences between the random and distinct bit selection strategies (Figure~\ref{fig:overhead}), the random strategy seems to be of higher practical relevance.

\subparagraph{Insertions.}
For insertions (Figure~\ref{fig:times}, column A), the throughput of Blocked Bloom filters and BlowChoc filters is higher than that of standard Bloom filters, except for small~$k\le 3$, where standard Bloom filters are competitive with Blocked Bloom filters, and BlowChoc filters perform poorly in comparison.
As expected, the Blocked Bloom filter has the highest throughput in most scenarios, followed by Blow2Choc and then by Blow3Choc filters.

As an example, consider $k=6$ with 9 threads (Figure~\ref{fig:times}(A) left at $k=6$): 
The Blocked Bloom filter's throughput is not 6 times as high as the standard Bloom filter's throughput, but only $1.6$ times, and while the gap widens for larger~$k$, it never reaches a factor of~$k$, since cache misses are not the only time-consuming operation for insertions, and our standard Bloom filter implementation benefits from prefetching the precomputed bit locations.
The throughput gap between 1 choice (Blocked) and 2 choices is larger than the one between 2 and 3 choices, which is explained by the introduction of a cost function for two or more choices, complicating the evaluation.

Parallelization is more effective for Blocked Bloom and BlowChoc filters than for standard Bloom filters, up to the point where all cores of the system are in use.
The poorer scaling of standard Bloom filters may be explained by the overall higher traffic on the memory bus caused by the random memory access pattern.
This also explains the steep decrease of throughput for standard Bloom filters between $3 \le k \le 10$.

\subparagraph{Lookups.}
For lookup throughput (Figure~\ref{fig:times}, columns B and C), we only consider the random bit selection strategy, as the distinct strategy is clearly not competitive.

We observe a clear advantage for Blocked Bloom filters, which only examine a single cache line.
BlowChoc filters are slower, but (with random bit selection) still outperform standard Bloom filters for successful lookups (Figure~\ref{fig:times}B), where all bits must be checked, whereas they show similar throughput for unsuccessful lookups.
Furthermore, the throughput is relatively independent of~$k$ for the Blocked Bloom and BlowChoc filters for $k\ge 10$, but for standard Bloom filters, this is only true for unsuccessful lookups, and we observe decreasing throughput for successful lookups.
Overall, unsuccessful lookups always show slightly higher throughput than successful lookups.
One reason for these observations is that for unsuccessful lookups at load $1/2$, only 2~bits need to be examined on average until a 0-bit is found.
This leads to large savings (in terms of cache misses) for Standard Bloom filters, but to lower savings (in terms of fewer local bit examinations) for Blocked Bloom filters with and without choices.
More precisely, the number of cache misses is the same for successful and unsuccessful lookups (1 for Blocked, 2 for Blow2Choc and 3 for Blow3Choc), but we have to examine more bits for successful queries ($k$ vs $\approx 2$). 

Parallelization across threads scales well with only slightly diminishing speedups as the number of threads increases.

\subsection{Application Cases in Genomics}
\label{sec:usecase}

Given the characteristics of standard and Blocked Bloom filters and BlowChoc filters, the following picture emerges:
For small $k\le 3$ (FPR $\ge 1/8$), standard Bloom filters can be used.
If speed is paramount and $3\le k\le 7$, or if memory usage is of no concern in case of larger~$k$, Blocked Bloom filters are the method of choice.
For larger $k > 7$, if available memory is limited, Blow2Choc filters offer a good compromise, requiring memory comparable to standard Bloom filters, but offering higher throughput.

Considering in addition the other filter types permitting online insertions (Cuckoo, Quotient, Vector Quotient filters; not evaluated here), one may ask whether there are realistic real application cases for BlowChoc filters.
We see such a scenario in genomics, where we represent either a reference genome (DNA sequence, a string over alphabet \texttt{ACGT}) or an individual sequenced sample as billions of distinct short fragments of DNA, each being $q$~nucleotides long, often $21\le q\le 31$.
These so-called $q$-grams can be encoded as 64-bit integers and stored as an approximate set using a probabilistic filter.
Clearly, the space consumption (for a given desired FPR) matters.
The Bloom filter family is useful because the space overhead factor is predictable (near 1.44, higher for the Blocked Bloom filter).
The non-Bloom filter types may have both lower and higher overhead factors (cf. Figure~\ref{fig:filtertypes}), depending on circumstances, making them more uncertain candidates in automated workflows.

As an example, we store several species' reference genomes as 31-gram sets in Bloom-type filters: the fruit fly \textit{Drosophila melanogaster} (122 million keys), the human \textit{Homo sapiens} (2.5 billion keys) and the axolotl \textit{Ambystoma mexicanum} (17.7 billion keys).
Table~\ref{tab:real_data} shows wall-clock build times and required space when using $k=14$ for an FPR of $2^{-14}$ with 9~inserter threads.
While the Blocked Bloom filter is always fastest and the Blow3Choc filter always requires the least memory, the Blow2Choc filter is a good compromise:
For the large genomes, the build time is roughly 1.5 times that of the Blocked Bloom filter, but only half of that of the standard Bloom filter, while using memory comparable to standard Bloom and Blow3Choc filters.
This can make a difference if one has, say, a system with 48~GB of RAM and wants to work on the axolotl genome.

As another example, recent work on xenograft sorting~\cite{xengsort}, concretely separating DNA reads from a mixture of mouse tissue and human tumor cells (engrafted in mouse tissue) may benefit from Blow2Choc filters by replacing the exact human/mouse $q$-gram representations with approximate ones with an FPR of $2^{-16} \approx 1/65\,000$.
The size would reduce for $q=27$ from the reported 17.3~GB to 13.3~GB, enabling the use of the application on a 16~GB multi-core laptop.
Detailed trade-offs would have to be evaluated in future work.

\begin{table*}[tb]\centering
\caption{
Wall-clock build times (9 inserter threads, minimum of 3 runs) and required filter sizes to represent different species' genomes as DNA 31-gram sets with an FPR of $2^{-14}$.
DNA data sources and genome statistics are given in Appendix~\ref{app:species}.
Best (lowest) values per column are marked in bold.
}\label{tab:real_data}
\addtolength{\tabcolsep}{-0.13em}
\begin{tabular}{l|cc|cc|cc|cc}
\hfill Species& \multicolumn{2}{c|}{Fruit fly (122M keys)} & \multicolumn{2}{c|}{Human (2.5B keys)} & \multicolumn{2}{c|}{Axolotl (17.7B keys)} \\ 
Filter type & Time [m:s] & Space [GB] & Time [m:s] & Space [GB] & Time [m:s] & Space [GB]\\
\hline 
standard Bloom & 0:08  & \textbf{0.53} & 2:17 & 6.42 & 25:55 & 43.93 \\
Blocked Bloom & \textbf{0:05} & 0.58 & \textbf{0:48} & 7.45 & \textbf{08:32} & 51.23\\
Blow2Choc & 0:06 & \textbf{0.53} & 1:12 & 6.48 & 12:25 & 44.38\\
Blow3Choc & 0:07 & \textbf{0.53} &  1:21 & \textbf{6.40} & 13:44 & \textbf{43.75} \\
\end{tabular}
\end{table*}

Furthermore, in genomics applications, the exact number of $q$-grams to be stored is usually not known in advance, although rough estimates can be given, e.g., we know that the human and comparable mammalian genomes have approximately $2.5$ billion $q$-grams for $q$ in a reasonable range around 31.
In such a situation, the Bloom filter family has advantages over the other filter types in Figure~\ref{fig:filtertypes}, which would fail if dimensioned too small, whereas the Bloom-type filters can be slightly overloaded at the cost of slightly increasing FPRs.


\section{Discussion and Conclusion}
\label{sec:discussion}

We have introduced Blocked Bloom filters with choices (BlowChoc filters).
They partially share the advantages of Blocked Bloom filters (cache-friendly fast insert and lookup), but avoid their most prominent disadvantage (larger space requirements) at low FPRs $2^{-k}$ for $k > 7$.
With 3 choices, BlowChoc filters can even reduce the memory requirements in comparison to Bloom filters for $k\ge 10$, but their lookup throughput reduces to that of standard Bloom filters.
Although Blow2Choc filters are slightly slower than Blocked Bloom filters and require slightly more space than Blow3Choc filters, they are a good compromise because they are still faster than standard Bloom and Blow3Choc filters and have an overhead factor comparable to standard Bloom filters for $k \in [10,20]$.
For $k \le 7$, Blocked Bloom filters provide a better time-memory trade-off compared to BlowChoc filters due to their fast insertion and lookup times.

Possibly, BlowChoc filters can be further improved by finding a better cost function for insertions, or by finding a faster distinct bit selection strategy.
We attempted a few other ideas (cost functions with more free parameters; using $k$~disjoint sub-blocks within a cache line), but none of them led to better FPRs.

We see good use cases for BlowChoc filters in genomics as a replacement for the commonly used (Blocked) Bloom filters or for larger exact representations of DNA $q$-grams.
Their online insertion capabilities and the possibility for slight overloading are essential properties not offered by other filter types.
The advantages of BlowChoc filters over standard or Blocked Bloom filters are most apparent for low FPRs, i.e., $k\ge 10$.

\bibliography{blowchoc}


\newpage\appendix
\part*{\sffamily Appendix}

\section{Cost Functions for Insertion}
\label{app:costfun}

We here describe several cost function types that we evaluated before choosing the exponential cost function used in the main article.

\subparagraph{Cost function arguments.}
All reasonable cost functions we invented are based on the following properties of a block:
\begin{itemize}
    \item $j$, the \emph{total} number of bits that would be set in the block \emph{after} insertion,
    \item $a$, the number of bits that would have to be newly set in the block.
\end{itemize}
In other words, if $F(x)$ is the set of bit addresses for a key $x$, and $F'$ is the set of already used bits in the block under evaluation, then $j=|F(x)\cup F'|$ and $a=|F(x) \setminus F'| = j - |F'|$.
Both values can be computed by simulating the insertion into the block and performing fast population count (\texttt{popcount}) operations on the 512 bits before and after insertion.

\subparagraph{Mixed cost based on FPR.}
The cost of inserting a key into a block with computed properties $(j,a)$ is based on the resulting local FPR $(j/512)^k$ in relation to the target FPR $(1/2)^k$, and on the number of additional bits~$a$ using a weight parameter $\sigma \ge 0$ as follows:
\begin{equation}\label{eq:cost:mix}
  \costsigma_{\sigma}(j, a) := \sigma k \cdot (j/512)^k / (1/2)^k + a = \sigma k \cdot (j/256)^k + a \,.
\end{equation}  
The choice of parameter~$\sigma$ has a considerable effect on the performance.
Intuitively, in an almost empty filter, the first term (resulting local FPR after insertion) is negligible, and the decision depends on where more bits can be re-used (lower $a$), 
However, if the target load of $1/2$ is reached with an insertion (and $\sigma=1$), then the FPR-based term becomes exactly $k$, which is also the maximally possible value of the second term~$a$.

\subparagraph{Lookahead cost.}
After moderate success with cost function~\ref{eq:cost:mix} due to some overfull blocks, we decided to design another cost function that penalizes already less full blocks stronger, by looking at the projected FPR cost after several ($\mu$) further worst-case insertions.
We define the lookahead cost
\begin{equation}
    \costmu_\mu(j,a) := \costsigma_1(j+\mu k, a) \,=\, k \cdot ((j + \mu k) / 256)^k \,+\, a \,.
    \label{eq:cost:la}
\end{equation}
This has the advantage that the cost for the local FPR dominates the cost for the newly set bits already at a lower load, which ideally should keep the overall load well below $1/2$.

\subparagraph{Exponential cost.}
After again moderate success, we decided to penalize blocks at risk of being overloaded even more strongly, and let the first term not model the FPR, but to increase it exponentially with the load with some base $\beta>1$.
We define the exponential cost
\begin{equation}
    \costbeta_\beta(j,a) := \beta^{(j / 128)} \,+\,  a / k \,.
    \label{eq:cost:exp:again}
\end{equation}
For a load of $1/4$ (128 set bits), the cost is $\beta$, and the cost for additionally set bits (relative to $k$) is added.

We evaluate the cost functions systematically in Appendix~\ref{app:bestcost} together with different bit selection strategies.

\section{Computing Distinct Bit Addresses}
\label{app:distinct}

\begin{figure}[!t]\centering
\vspace*{4ex}
\includegraphics[width=0.9\linewidth]{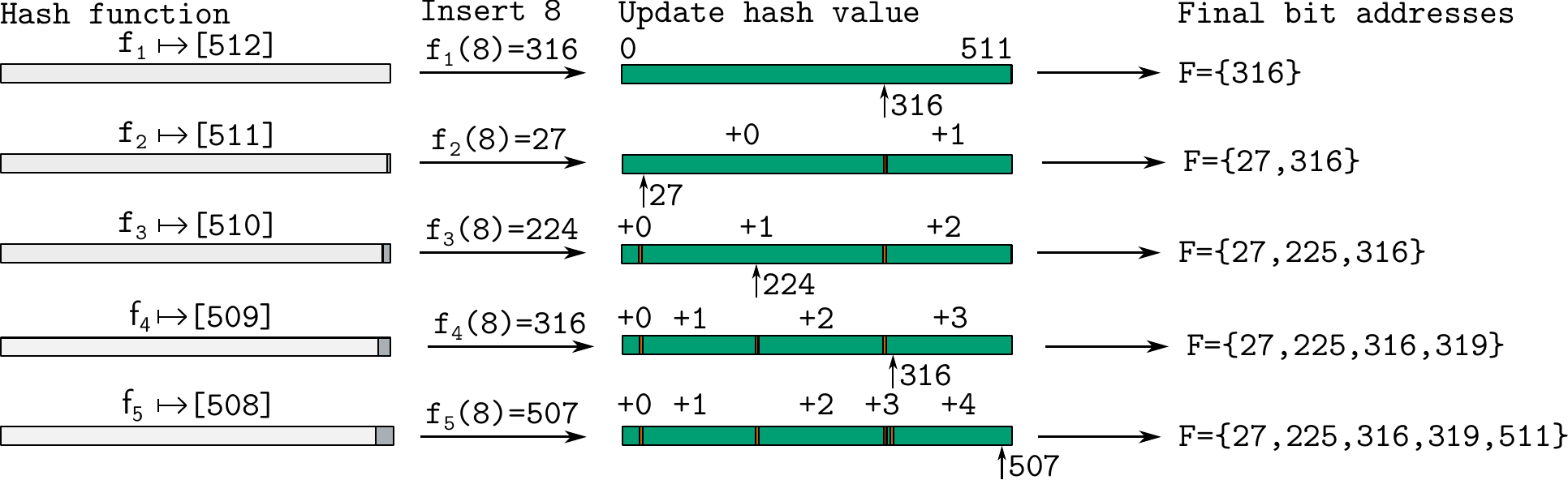}
\caption{
We compute $k=5$ distinct bit addresses for key $x=8$.
The first $f_1: \universe\to [512]$ maps to all possible positions in a block $\{0,\ldots,511\}$ (indicated by the light gray bar one the left side).
The bit position $f_1=316$ is marked with an arrow in the green block which visualizes the already used positions. 
The resulting set of distinct hash positions is $F=\{316\}$ (on the right).
For the second hash function $f_2: \universe\to [511]$, only $511$ positions are left. 
This is indicated by the dark gray part of the bar on the left side.
The green bar of used positions can be divided into two parts. 
In the first part, in which $f_2(8)=27$ is located, the value remains unmodified.
In the second part, we add an offset of 1 for all values $\geq 316$. 
This is necessary since 316 is already used and excluded from the possible positions.
For the following hash functions, the hash value is offset by the number of occupied slots which are smaller or equal to the new bit address to insert into $F$.
}\label{fig:distinct}
\end{figure}

To compute $k$~distinct bit addresses for a key~$x$, we first compute hash values $f_i(x)$ for $i=1,\dots, k$ as follows:
Each hash function $f_i$, is chosen randomly and independently from some universal family and is assumed to represent an independent random variable taking uniformly distributed values in the set $[512-i+1]$.
Before we insert $f_i(x)$ into the set of bit addresses $F$, we increment $f_i(x)$ as long as we have values in $F$ (for hash functions $f_j, j < i$) that are smaller or equal to the updated value of $f_i(x)$.
During this procedure, we always keep the $f_j(x)$,  $j\le i$, in sorted order.
Since each $f_i$ maps to a smaller range, it is guaranteed that the final value of $f_i(x)$ is in $[512]$ when it is added to the set $F$ and that the values are distinct.
Pictorially speaking, by choosing an initial random value in $[512-i+1]$ for $f_i$, it in fact picks a random value from $[512] \setminus \{f_1(x), \dots, f_{i-1}(x)\}$.
A detailed example is given in Figure~\ref{fig:distinct}.
The implementation uses a cache-friendly dynamic insertion sort for every computed and adjusted $f_i(x)$.
Even though these are fast localized operations, we observe significant computational overhead compared to random bit selection.

\section{Evaluation of Cost Functions}
\label{app:bestcost}

\begin{figure*}[t]\centering
\includegraphics[width=1.0\linewidth]{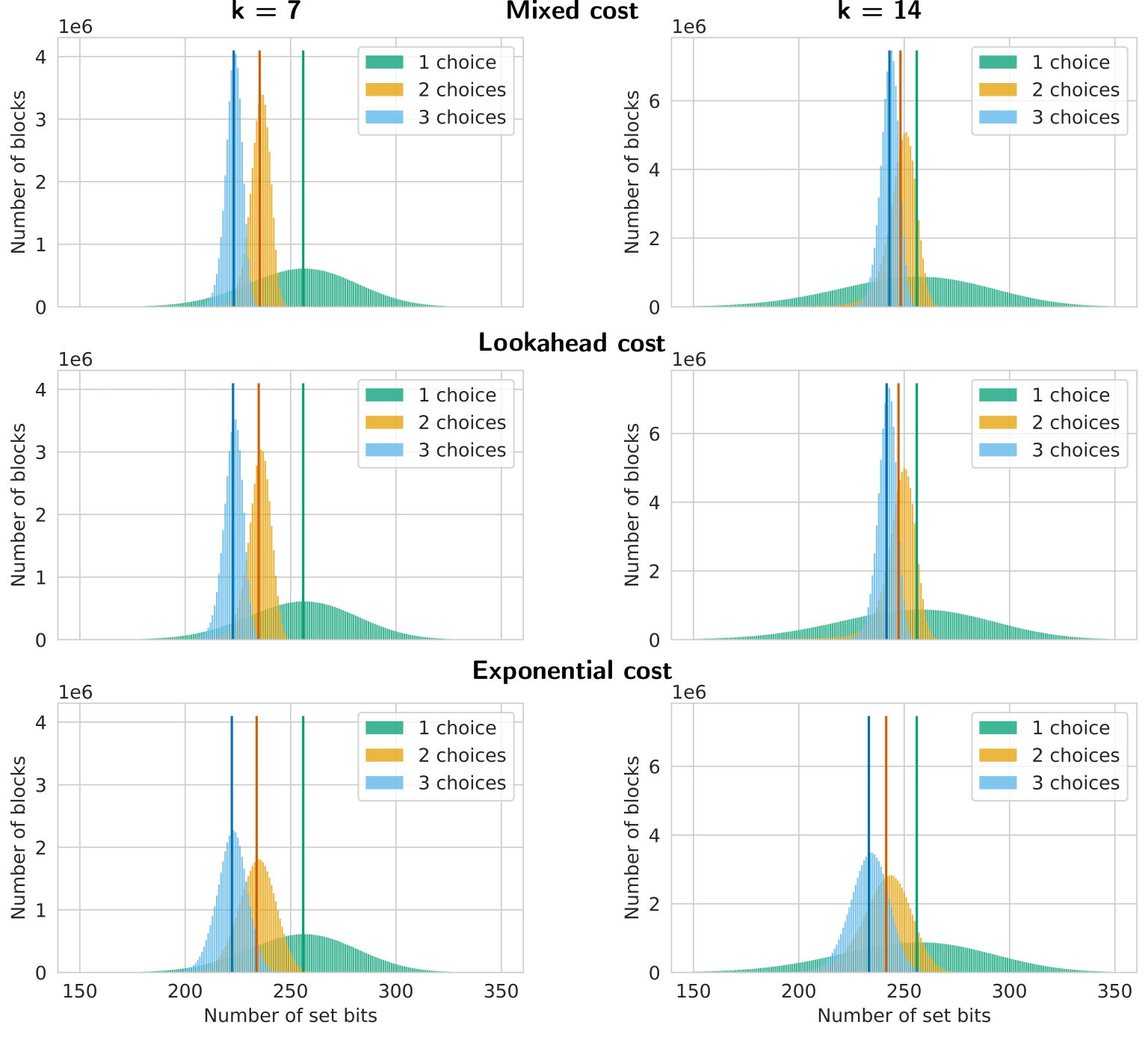}
\caption{
Histogram of the number of set bits in each block after insertion using ``random'' bit selection for $k=7$ or $k=14$ bit address hash functions (left column: $k=7$; right column: $k=14$), different cost functions (top row: mixed cost with $\sigma = 2.5$ for $k=7$ and $\sigma = 50$ for $k=14$; middle row: lookahead cost with $\mu=3.5$); bottom row: exponential cost with $\beta = (1 + \sqrt{5})/2$), and different choices ($c=1,2,3$; color).
Vertical lines of the respective colors show the distributions' means.
}\label{fig:histogram}    
\end{figure*}

We evaluate BlowChoc filters using the three different cost types defined in Appendix~\ref{app:costfun} ($\costsigma, \costmu, \costbeta$).
For a given target FPR of $\eps=2^{-k}$, we select the filter size $m$ (in bits) for the standard Bloom filter ($m = nk/\ln(2)$, rounded up to the next multiple of 512) and use a block size of $M := m / 512$.
For a standard Bloom filter, this results in an exact FPR of $2^{-k}$.
However, the FPR of a Blocked Bloom filter (without choices) is worse (and increases with~$k$, see~Section~\ref{sec:compare:size}).
For a BlowChoc filter, the resulting FPR depends on the number of choices, the cost function and its parameters, and the bit selection strategy (random or distinct).
With two choices, the FPR is \emph{per se} (almost) twice as high as for a Blocked Bloom filter without choices, but this effect is compensated by a lower overall average load and additionally by better load balancing, i.e. a smaller variance of the load across blocks.
This effect is even more noticeable for three choices, with a \emph{per se} tripled FPR.
In Figure~\ref{fig:histogram}, we observe that BlowChoc filters with three choices achieve a lower average and more balanced load distribution than for two choices, which again show a lower average and more balanced distribution than the standard Blocked Bloom filters (1 choice).

However, in order to achieve comparable FPRs, the average load of Blocked Bloom filters with choices must be reduced considerably, especially for small~$k$ (Figure~\ref{fig:maxbits}).
For example, for $k=7$ the average number of set bits per block must decrease from 256 to $\le 232$ bits for two choices and to $\le 219$ bits for three choices. 
For larger~$k$, the required difference of set bits per block for an FPR of $2^{-k}$ is smaller between two and three choices (e.g.\ 244 vs.\ 237 for $k=14$), and it can hence be achieved more easily.

\begin{figure*}[bp]\centering
\includegraphics[width=1.0\linewidth]{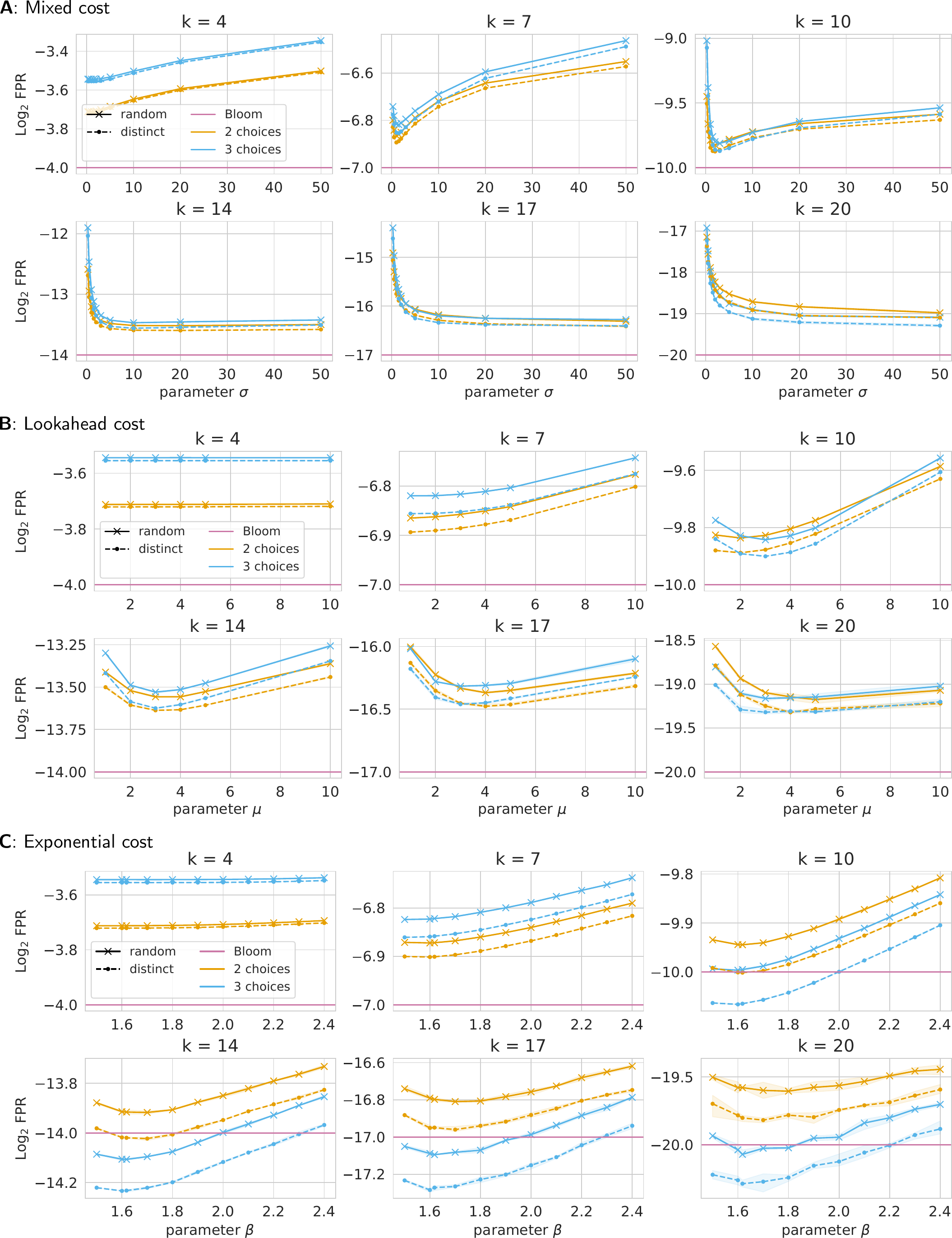}
\caption{
Empirical $\log_2$\,FPR of BlowChoc filters for three different cost types (panels \textbf{A, B} and \textbf{C}) and their respective $\sigma$, $\mu$ and $\beta$, for different numbers of choices (color: 2, 3), bit selection strategies (line styles; random: solid , distinct: dashed), and number of hash functions $k$ (subfigures within a panel).
Lines show averages and shaded areas (mostly visible for $k=20$) show ranges between minimum and maximum observed FPR of three runs.
The horizontal line for standard Bloom filters is at the targeted FPR of $2^{-k}$.
Blocked Bloom Filter FPRs without choices are not shown for readability; they are higher than for 2 or 3 choices and would be off scale.
}\label{fig:cost_functions}
\end{figure*}

Figure~\ref{fig:cost_functions} shows the resulting $\log_2$ FPR for $k\in \{4,7,10,14,17,20\}$, different bit selection strategies (random, distinct), choices (2, 3) and cost functions.
Various interesting observations can be made.

For all cost functions, values of $k$, choices and parameters, the distinct bit selection strategy has a smaller FPR compared to the random bit selection strategy, as expected.
The optimal cost parameter ($\sigma$ for $\costsigma$, $\mu$ for $\costmu$, and $\beta$ for $\costbeta$) depends on $k$ and the number of choices.
Using the mixed or lookahead cost functions, we are not able to achieve the target FPR of the standard Bloom filter with any parameter values.
The exponential cost function ($\costbeta$) leads to the overall best FPRs and with a surprisingly consistently good parameter value of $\beta\approx 1.6$ or the golden ratio $\beta \approx (1 + \sqrt{5})/2 \approx 1.618$ for all choices, bit selection strategies, and $k$.
In the main evaluations, we only consider $\costbeta_\beta$ with $\beta = (1 + \sqrt{5})/2$.
The fact that close-to-optimal values for $\beta$ are close to the golden ratio appears to be purely coincidental.

Two choices lead to lower FPRs than three choices for $k < 10$.
In fact, for $k=4$, the Blocked Bloom filter without choices is even better (cf.\ Figure~\ref{fig:overhead}).
However, for $k \geq 10$, the BlowChoc filters with three choices reach a lower FPR compared to two choices or Blocked Bloom filters.
For three choices, $k\ge 10$ and both bit selection strategies, BlowChoc filters may even achieve the same or a better FPR than standard Bloom filters, which Blocked Bloom filters never do.
Alternatively, the filter may be slightly smaller to precisely achieve the target FPR of $2^{-k}$.

\section{Genome Data Sources and Statistics}
\label{app:species}

The fruit fly \textit{Drosophila melanogaster} has a relatively small genome with 142 million base pairs of DNA and 122 million distinct 31-grams.
We used the following reference provided by NCBI:
\url{https://ftp.ncbi.nlm.nih.gov/genomes/all/GCF/000/001/215/GCF_000001215.4_Release_6_plus_ISO1_MT/GCF_000001215.4_Release_6_plus_ISO1_MT_genomic.fna.gz}

The human \textit{Homo sapiens} genome has 3.1 billion base pairs of DNA and 2.5 billion distinct 31-grams.
We used the following reference provided by the T2T consortium \cite{t2t}:
\url{https://s3-us-west-2.amazonaws.com/human-pangenomics/T2T/CHM13/assemblies/analysis_set/chm13v2.0.fa.gz}

The axolotl \textit{Ambystoma mexicanum} is one of the species with the largest known genomes. 
Its genome contains 29.1 billion base pairs and 17.7 billion distinct 31-grams.
We used the following reference provided by NCBI:
\url{https://ftp.ncbi.nlm.nih.gov/genomes/all/GCF/040/938/575/GCF_040938575.1_UKY_AmexF1_1/GCF_040938575.1_UKY_AmexF1_1_genomic.fna.gz}

The download links may also be found in the \texttt{benchmark\_species} workflow in the accompanying implementation (see abstract).
The reference genomes are automatically downloaded when the workflow is executed.


\end{document}